\documentclass{elsart}
\usepackage[dvips]{graphicx}

\usepackage{amssymb}

\begin{document}

\begin{frontmatter}



\title{Geophysical aspects of very long baseline neutrino experiments}


\author{Robert J.\ Geller} 
\address{Department of Earth and Planetary Science,
Graduate School of Science, Tokyo University, 
Hongo 7-3-1, Bunkyo-ku, Tokyo 113-0033, Japan}
\ead{bob@eps.s.u-tokyo.ac.jp} 

\author{Tatsuhiko Hara} 
\address{International Institute of Seismology and
Earthquake Engineering, Building Research Institute, 
Tatehara 1, Tsukuba 305-0802, Japan}
\ead{thara@kenken.go.jp}


\begin{abstract}
Several proposed experiments will send beams of neutrinos through the Earth along
paths with a source-receiver distance of hundreds or thousands of
kilometers.   Knowledge of the physical properties of the 
medium traversed by these beams, in particular the density, will be
necessary  in order to properly interpret the experimental data.   
Present geophysical knowledge allows the average density along a path 
with a length 
of several thousand km to be estimated with an accuracy of 
about $\pm 5$~per cent.
Physicists planning neutrino beam experiments should decide whether or not this
level of uncertainty is acceptable.  If greater accuracy is required,
intensive geophysical research
on the Earth structure along the beam path should be conducted as 
part of the preparatory work on the experiments.  

\end{abstract}

\begin{keyword}
Long baseline neutrino experiments, Earth's density distribution  


\PACS 13.15.+g  14.60.Pq  23.40.Bw  91.35.-x

\end{keyword}

\end{frontmatter}


\section{Introduction}

The Earth is a tectonically active planet. 
Large scale thermal convection, which is 
related to the motion of tectonic plates on the Earth's surface, is 
taking place in the Earth's crust and mantle (which collectively 
extend from the core-mantle boundary, 
$r \approx 3480$~km, to the Earth's surface, $r \approx 6371$~km,
where $r$ is the radius\footnote{The Earth is ellipsoidal, due primarily to 
its rotation; except where otherwise noted, 
depths are spherically averaged values.}).   
The crust and mantle consist primarily of silicate rocks.
The outermost layer is the crust, which has a thickness ranging from
about 80~km under Tibet to about 5~km in beneath oceans.   
As discussed below, the physical
properties of the crust are highly laterally heterogeneous.
The mantle is divided into the upper mantle, which extends from the
base of the crust to a depth of about 410~km ($r \approx 5961$~km); the transition
zone, in the depth range 410~km to 660~km ($r \approx 5711$~km to $r  \approx  5961$~km);   
and the lower mantle, in the depth range 660~km to 2891~km ($r \approx  3480$~km
to $r \approx 5710$~km).   The boundaries between the upper mantle
and the transition zone, and between the transition zone 
and the lower mantle, are thought to be
due to phase transitions in silicate minerals.

The Earth's core consists primarily of
iron and thus has a considerably greater density than the
mantle.  The outer core 
($r \approx 1222$~km to $r \approx 3480$~km) is liquid; 
magnetohydrodynamic convection in the outer core is considered to be the
cause of the Earth's magnetic field.   The inner core, which extends 
from the base of the outer core to the Earth's center ($r=0$ 
to $r \approx 1222$~km), is solid.  

For further general information on the structure of the 
Earth's interior see recent
textbooks (e.g., Lay and Wallace, 1995; Shearer, 1999) 
and the works cited therein.  

Due to the increase of pressure with depth, the Earth's density and elastic constants
are vertically heterogeneous.  However, 
because the Earth is tectonically active, its physical properties are also
laterally heterogeneous.   Let us denote the laterally averaged 
one-dimensional (1-D) 
density structure by $\rho (r)$, where $\rho$ is the density in units
of gm/cm$^3$ (or kg/m$^3$), and denote 
the three dimensional (3-D) density distribution by $\rho (r, \theta , \phi)$,
where $\theta$ and $\phi$ are respectively the colatitude and longitude, 
in spherical polar coordinates. 

The Earth's average density can be determined from its total mass, 
$m_e = 5.97 \times 10^{24} $~kg, 
and its outer radius.  If the Earth were a homogeneous sphere, its moment of inertia
would be $0.40Mr_e^2$, where $r_e$ is the Earth's outer radius.
However, the observed moment of inertia has a much smaller value,
approximately $0.33Mr_e^2$.  This confirms that the Earth's inner regions (i.e.\
the outer and inner core) are significantly denser than average. 

Even if the Earth's total mass and moment of inertia are combined with 
other geodetic data such as the spherical harmonic
expansion of the Earth's external gravity field (which is inferred from satellite data),
these data provide integral constraints on the Earth's density distribution   
but are insufficient to determine it uniquely.  It therefore is necessary
to use seismological data as the primary basis for inferring the Earth's
density distribution.  However, 
for technical reasons that are not discussed in detail here, inferring 
the Earth's density distribution 
directly from observed seismological data 
is not practically realizable  
(Bullen, 1975; Kennett, 1998).  
It thus is 
necessary to follow a two step inference process.   First the
spatial distribution of seismic wave velocities is inferred
from seismological data;
second, the density distribution is inferred from the seismic velocities, using  
the above integral constraints together with other empirical relations.    
Both of these steps introduce uncertainty 
into the density model.

\subsection{Seismic velocities}

For the purposes of very long baseline 
neutrino experiments,  isotropic Earth models can probably be
regarded as sufficiently accurate; the discussion in this paper is
limited to such models.  
The most general anisotropic elastic solid has 
21 independent elastic constants, but 
an isotropic elastic solid has only two independent elastic constants, 
the Lam\'{e} constants $\lambda$ and $\mu$.  

In an isotropic elastic body the velocity of compressional 
elastic waves (P-waves), $\alpha$, and the velocity of transverse elastic waves
(S-waves), 
$\beta$, are given respectively by 
\begin{eqnarray}
\alpha & = & \sqrt{(\lambda + 2 \mu)/ \rho}  \\ \label{vp}
\beta  & = & \sqrt{ \mu / \rho} .   \label{vs}
\end{eqnarray}
As a rough approximation, the ratio of P- and S-wave velocities in the solid
Earth is given by
\begin{equation}
\alpha \approx 1.7 \beta ,    
\end{equation}
but the exact value of the proportionality constant varies 
with the chemical composition, pressure and temperature.

\section{How Earth models are inferred} 

Inversion of observed seismic data for Earth structure is 
an underdetermined inverse problem, 
and all Earth models are subject to error and uncertainty.
Regularization 
constraints of some type (e.g., smoothness, minimum variation from
the starting model, etc.) must be applied to obtain a stable 
solution.   Inverse theory
allows formal error estimates to be made, but it is well
known that systematic errors, which
cannot be quantitatively estimated, may often
be on the same order or larger.   Systematic errors are due
to factors such as the uneven distribution of seismic observatories
on the Earth's surface (in particular the lack of observatories on
the ocean bottom) and the uneven spatial distribution of earthquakes, and
thus cannot be easily reduced. The approximations 
(e.g., ray-theoretic, linearized perturbation with respect to a spherically
symmetric model, etc.) used to model seismic wave propagation
are another significant source of systematic
errors; progress in forward modeling and inversion techniques is
leading to reduction of such errors.  
Anelastic attenuation (absorption) of seismic waves 
also places inherent limits on resolving power,
especially of deeper and shorter wavelength structure.

A 1-D model seismic velocity specifies 
$\alpha (r)$ and $\beta (r)$, while a 3-D model specifies 
$\alpha (r, \theta, \phi )$ and $\beta (r, \theta, \phi )$.  
A 1-D model may either be a globally averaged model or a model
of the depth dependence under some region; similarly, a 3-D
model may either be a global model or may be limited to some
particular region.  
The main focus of seismological research on Earth structure
has shifted to the quest to infer 3-D Earth models.    
In this context, the role of 1-D models is to provide the
starting point for defining a 3-D model as a perturbation 
to the 1-D starting model.

The primary data used to obtain seismic velocity models are the arrival 
times of seismic body waves (P- and S-waves that travel through the 
Earth's interior).   The arrival time data 
are then analyzed to determine the location $(r_0, \theta_0 ,
\phi_0 )$ and origin time $t_0$ of each earthquake and  
can then be
converted to the travel time from the source to the receiver.    
A large dataset of travel time data for many earthquakes is then 
inverted to obtain a new Earth model, and the earthquake location
process is then updated.  This process is iterated   
several times until convergence is obtained.    
Travel time data are in some cases supplemented by surface wave dispersion
data (the frequency dependence of the phase and group velocities of 
seismic surface waves)
or free oscillation data (the frequencies of several hundred of the longest 
period modes, which are basically equivalent to surface waves). 
A recent trend is to use the seismic waveforms themselves (the recorded
displacement of the ground as a function of time), rather than
secondary data such as the travel times, as the data in the 
inversion. 

Improvements in data and in inversion methodology over the past
20 years have led to steady improvement in seismic velocity models. 
Two well known 1-D models are the 
``Preliminary Reference Earth Model'' (PREM)
of Dziewonski and Anderson (1981) and model ak135 (Kennett {\em et al.}, 1995).
The latter is based on a more extensive dataset than the former, 
and is therefore more accurate.
Research on 3-D Earth structure is a highly active field; recent reviews
by Garnero (2000) and Nataf (2000) provide a useful starting point.

Lateral variation of elastic properties and density is greatest in the 
crust and uppermost mantle, but the density of broadband seismic observatories used
for global seismology is far too small (especially in view of the
non-uniform geographical distribution) to determine the lateral 
heterogeneity of the ``crustal
structure'' (where this term includes both the crust and the 
uppermost mantle).   Geophysicists must
therefore use data collected from various local and regional surveys
to correct for the effect of crustal structure so that their
data can then be analyzed to determine 3-D Earth structure on a global
scale.  Two widely used models for this purpose are CRUST 5.1
(Mooney {\em et al}., 1998), which has a resolution of 5$^o$ (i.e., about
500~km), and its successor, Crust 2.0 
(http://mahi.ucsd.edu/Gabi/rem.dir/crust/crust2.html), 
with a resolution of 2$^o$.   
These models are not intended as accurate models
of the crust, but rather are intended as ``pretty good'' models, for the 
purpose of removing crustal effects. 
Physicists planning neutrino beam experiments should exercise
appropriate caution when using these models.

\subsection{Density models}

Both global and regional density models are subject to 
considerable uncertainty.
Global scale density models are typically derived by applying an 
equation of state, which is an empirical approximation, 
to seismic velocity models (e.g., Bullen, 1975).    Crustal 
density models are derived using a variety of empirical relations
between seismic velocities and densities (see Mooney {\em et al.}, 1998).
It is striking that, especially for the case of sedimentary rocks,
many of these empirical relations were published in the 1970s, which
suggests that there has not recently been a high level of activity in this field.

It is difficult to quantify the 
uncertainty of published density models.  
One interesting approach  
is that of Kennett (1998).  He exploited the fact that 
the frequencies of the longest period modes of the Earth's free
oscillations depend separately on the elastic constants and the density
to a marginally resolvable extent to 
conduct the following numerical experiment.  
He fixed the seismic velocities and density of his Earth model
to the values of the PREM model,
and constructed a random ensemble of density models centered around the
PREM model.   He then calculated the free oscillation eigenfrequencies
for each model and compared them to the observed eigenfrequencies
to construct a set of the 50 best fitting models.  These 
density models, all of which can be said to fit the 
free oscillation data acceptably, have a range 
of about $\pm 2 \sim 3$~per cent in the upper mantle.     
This should not be regarded as a
conclusive error estimate, but it is one reasonable indication of the
general level of uncertainty of present 1-D global density models.     

\section{Density models for neutrino beam experiments}

\begin{figure}
\begin{center}
\includegraphics*[width = 11 cm, trim = 0cm 6.3cm 0cm 0cm ]{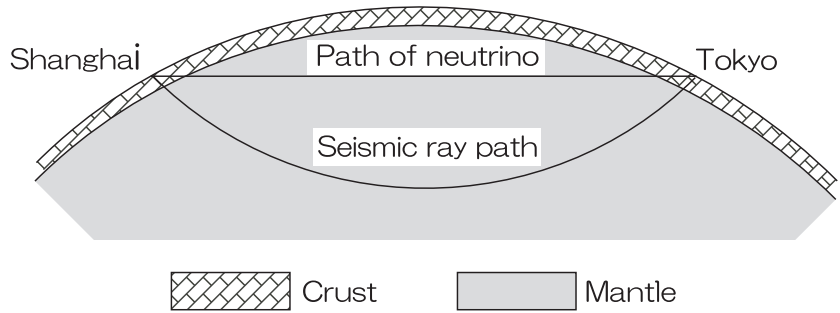}
\end{center}
\caption{Schematic depiction of a hypothetical neutrino beam
experiment.}
\label{Fig1}
\end{figure}

Let us consider a hypothetical neutrino beam
experiment (Fig.~\ref{Fig1})  with a neutrino source in Tokyo
and a detector in Shanghai.   Note that the neutrino beam
follows a straight line, but a seismic wave from Tokyo 
to Shanghai (or vice versa) follows a curved path
(the path of minimum travel time).   Thus it is not possible
to infer the physical properties of the
neutrino beam path based only on observations of seismic waves
traveling from Tokyo to Shanghai.
Published 3-D Earth models, 
which were obtained by analyzing a large dataset using many
sources and receivers, can be used to obtain a seismic velocity
profile along the neutrino beam path, which can then be empirically
converted to density.   If the accuracy of the density profile
obtained using the above procedure is deemed insufficient, 
further information could in principle be obtained by conducting 
a seismic observation 
campaign with receivers along the entire great circle 
from Tokyo to Shanghai.  However, the fact that 
much of the beam path lies under the oceans would greatly 
complicate such a campaign. 

\begin{figure}
\begin{center}
\includegraphics*[width = 11 cm]{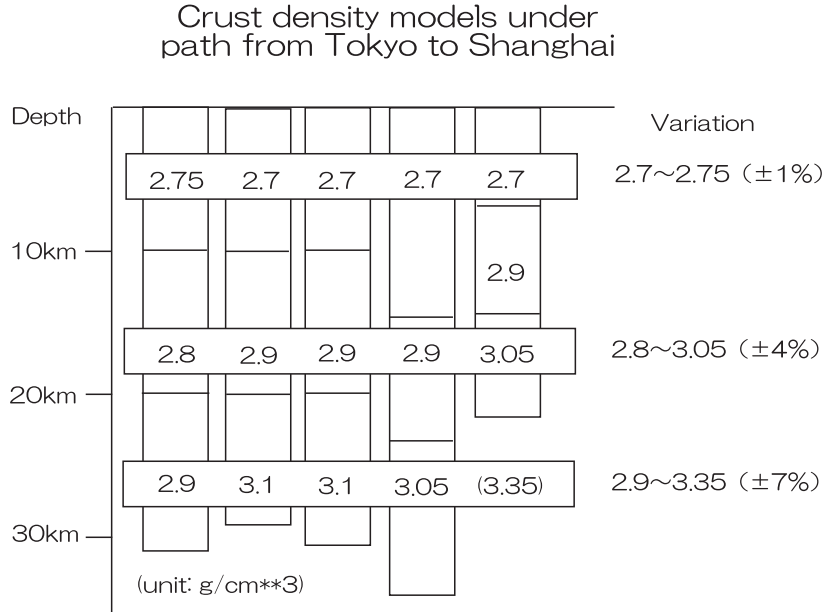}
\end{center}
\caption{The various density profiles under the
Tokyo-Shanghai path, from model Crust 2.0.   Order from left to right
is arbitrary.}
\label{Fig2}
\end{figure}

Figure~\ref{Fig2} shows the various density profiles 
under the hypothetical Tokyo-Shanghai path, taken from
Model Crust 2.0.  As shown in
Fig.~\ref{Fig2}, the variation between the various density
profiles is $\pm 4$~per cent in the depth range from
10--20~km
and $\pm 7$~per cent in the depth range from 20--30~km.
The variations in density are due to the differences in the 
physical properties of the various types of geological
units, but can also be regarded as a crude indicator of
the general level of uncertainty of the density.   
As, generally
speaking, the amplitude of the Earth's lateral
hetereogeneity decreases with increasing depth, the
variability of $\pm 7$~per cent in Fig.~\ref{Fig2} can reasonably 
be regarded as as an upper bound on the uncertainty.    
Note that the density in the depth range 20--30~km in
the rightmost column of  Fig.~\ref{Fig2} (3.35~g/cm$^3$)
is the value for the uppermost mantle, and is about
10~per cent higher than the density of the lowermost
crust.   

\section{Discussion}

Neutrino beam physicists should be aware of the various
uncertainties and limitations of present geophysical knowledge
of the Earth's density distribution, as discussed in this paper.
The planning of neutrino beam experiments should
include simulation of the data reduction process, including a
propagation of error analysis, to study the effect of this uncertainty.  
Three possible scenarios
can be envisioned.  (1) The uncertainty of present density models
poses no significant problems; (2) moderate reduction of the 
uncertainty, through more detailed analysis of existing data, is required:
(3) significant reduction of this uncertainty, 
by conducting a large scale campaign of geophysical observations, is required.
Obviously, scenario (1) would be most desirable, while scenario (3) would
be discouraging.  This issue should be resolved 
at an early stage of the planning of neutrino beam 
experiments.

\section*{References}

\noindent
Bullen, K., 
{\em The Earth's Density}, Chapman \& Hall (London, 1975).  

\noindent
Dziewonski, A.\ M., \&  Anderson, D.\ L., Preliminary
reference Earth model, 1981, Phys.\ Earth Planet.\ Int., 
25, 297-356.

\noindent
Garnero, E.\ J., Heterogeneity of the lowermost mantle, 2000, 
Ann.\ Rev.\ Earth Planet.\ Sci., 28, 509-537.  

\noindent
Kennett, B.\ L.\ N.,  On the density distribution
within the Earth, 1998, Geophys.\ J.\ Int.,
132, 374-382.   

\noindent
Kennett, B.\ L.\ N., Engdahl, E.\ R., \& Buland, R., 
Constraints on seismic velocities in the Earth from travel-times, 1995, 
Geophys.\ J.\ Int., 122, 108-124.  

\noindent
Lay, T., and T.\ C.\ Wallace, {\em Modern Global Seismology},
Academic (San Diego, 1995).  

\noindent
Mooney, W.\ D., Laske, G., \& Masters, T.\ G.,
Crust 5.1: A global crustal model at $5^o \times 5^o$, 1998, 
J.\ Geophys.\ Res.,  103, 727-747.

\noindent
Nataf, H.-C., Seismic imaging of mantle plumes, 2000, 
Ann.\ Rev.\ Earth Planet.\ Sci., 28, 391-417.

\noindent
Shearer, P.\ M., {\em Introduction to Seismology},
Cambridge U.\ Press (Cambridge, 1999).








\end{document}